\documentclass[aps,pre,twocolumn,showpacs,10pt]{revtex4-1}

\usepackage{graphicx} 
\usepackage{color}
\definecolor{Green}{rgb}{0,0.5,0}

\usepackage[english]{babel}
\usepackage[utf8]{inputenc}

\usepackage[babel=true]{csquotes}

\usepackage{amssymb}
\usepackage{amsmath}

\allowdisplaybreaks

\usepackage{textcomp}

\newcommand{\myUnit}[1]{\,\mathrm{#1}}

\usepackage{refstyle}

\newref{fig}{%
  name   = fig.~,
  names  = figs.~,
  Name   = Fig.~,
  Names  = Figs.~,
  rngtxt = \RSrngtxt,
  lsttxt = \RSlsttxt,
}

\newref{figs}{%
  name   = figs.~,
  names  = figs.~,
  Name   = Figs.~,
  Names  = Figs.~,
  rngtxt = \RSrngtxt,
  lsttxt = \RSlsttxt,
}

\newref{sec}{%
  name   = sec.~,
  names  = secs.~,
  Name   = Sec.~,
  Names  = Secs.~,
  rngtxt = \RSrngtxt,
  lsttxt = \RSlsttxt,
}

\newref{chp}{%
  name   = chap.~,
  names  = chaps.~,
  Name   = Chap.~,
  Names  = Chaps.~,
  rngtxt = \RSrngtxt,
  lsttxt = \RSlsttxt,
}

\newref{eq}{%
  name   = eq.~,
  names  = eqs.~,
  Name   = Eq.~,
  Names  = Eqs.~,
  rngtxt = \RSrngtxt,
  lsttxt = \RSlsttxt,
}

\newref{tab}{%
  name   = tab.~,
  names  = tabs.~,
  Name   = Tab.~,
  Names  = Tabs.~,
  rngtxt = \RSrngtxt,
  lsttxt = \RSlsttxt,
}

\newref{app}{%
  name   = app.~,
  names  = apps.~,
  Name   = App.~,
  Names  = Apps.~,
  rngtxt = \RSrngtxt,
  lsttxt = \RSlsttxt,
}

\begin{document}

\title{Measuring the local quantum capacitance of graphene using a strongly coupled graphene nanoribbon}
\author{D. Bischoff}
\email{dominikb@phys.ethz.ch}
\author{M. Eich}
\author{A. Varlet}
\author{P. Simonet}
\author{T. Ihn}
\author{K. Ensslin}
\affiliation{Solid State Physics Laboratory, ETH Zurich, 8093 Zurich, Switzerland}
\date{\today}



\begin{abstract}

We present electrical transport measurements of a van-der-Waals heterostructure consisting of a graphene nanoribbon separated by a thin boron nitride layer from a micron-sized graphene sheet. The interplay between the two layers is discussed in terms of screening or, alternatively, quantum capacitance. The ribbon can be tuned into the transport gap by applying gate voltages. Multiple sites of localized charge leading to Coulomb blockade are observed in agreement with previous experiments. Due to the strong capacitive coupling between the ribbon and the graphene top layer sheet, the evolution of the Coulomb blockade peaks in gate voltages can be used to obtain the local density of states and therefore the quantum capacitance of the graphene top layer. Spatially varying density and doping are found which are attributed to a spatial variation of the dielectric due to fabrication imperfections.
\end{abstract}
%


\maketitle


One of the advantages of layered materials such as graphene, hexagonal boron nitride (hBN) or various dichalcogenides, is the possibility to fabricate stacks in order to obtain a heterostructure material with new properties. This was recently exploited to fabricate a variety of different device structures~\cite{Georgiou2012,Britnell2012science,Britnell2012nano,Yang2012,Choi2013,Lee2013,Lee2014,Liu2012} and to investigate a number of novel effects: localization in super clean systems~\cite{Ponomarenko2011,Kechedzhi2012}, current drag~\cite{Gorbachev2012,Kim2012,Titov2013} and the emergence of superlattices~\cite{Yankowitz2012,Dean2013,Ponomarenko2013}. Most of these experiments were performed with so-called \enquote{double-layer graphene}, consisting of two graphene layers separated by a thin insulator layer. In contrast to Bernal-stacked bilayer graphene, charge transfer between the two layers is not possible. 

Above experiments were conducted with micron-sized graphene sheets. In this paper we develop this idea one step further: we pattern the lower graphene sheet laterally into a nanoribbon (e.g. Refs.~\cite{Han2007,Todd2009,Molitor2009,Liu2009,Bischoff2012}) and stack an insulator plus a micron-sized graphene sheet on top. This experiment is a crucial step towards fabricating and understanding stacked graphene nanostructures. If the nanoribbon is tuned into the so-called transport gap~\cite{Todd2009,Molitor2009}, Coulomb blockade can be observed~\cite{Todd2009,Molitor2009}. The localized charge in the ribbon serves as a sensitive and local detector for its electrostatic surrounding and allows us to probe properties of the graphene top layer. This is conceptually similar to an SET above a 2DEG~\cite{Wei1998} and scanning SET measurements~\cite{Hess1998,Martin2009}. Despite using a comparably thick insulator layer ($13\myUnit{nm}$), we can still reach the regime where the spacing between the layers is smaller than the average distance between two charge carriers within one layer~\cite{Gorbachev2012}.

In this paper we present simultaneous electrical transport measurements through a graphene nanoribbon and a graphene sheet (top layer) that are separated by a layer of hBN, and therefore strongly capacitively coupled. By applying gate voltages, we can independently tune both the graphene top layer and the graphene ribbon to their respective Dirac point. If the graphene ribbon is tuned close to the Dirac point, electrical transport is dominated by Coulomb blockade due to localized charges. We use the evolution of Coulomb resonances in gate voltages to extract the local density and quantum capacitance of the graphene top layer.

\begin{figure}[tbp]
\centering
\includegraphics[width=1.0\columnwidth]{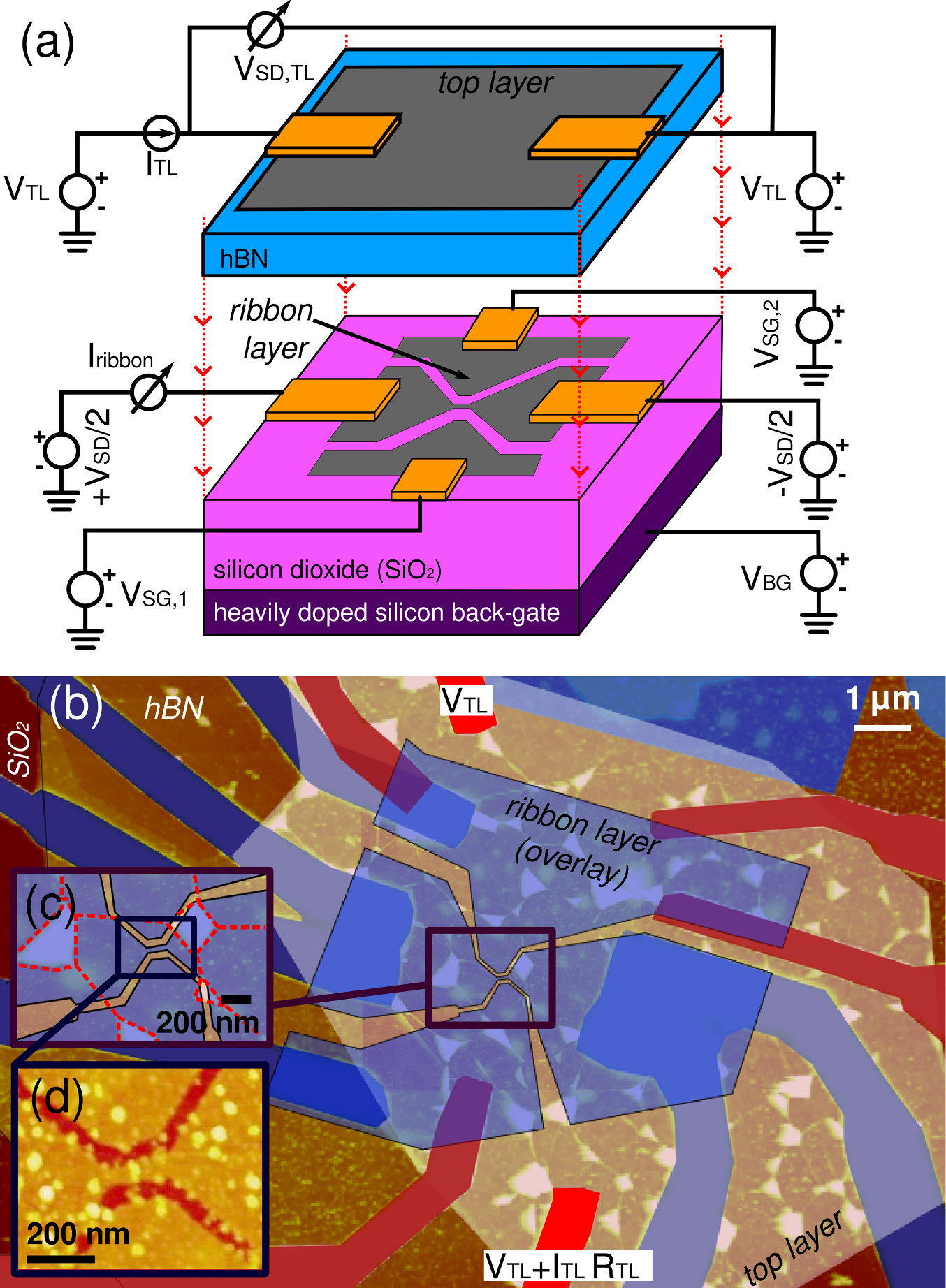} %
\caption{(color online) (a) Schematic of the device: a graphene ribbon is fabricated on a SiO$_2$ substrate. Source, drain and two side gates are contacted by gold contacts (ribbon layer). A hBN flake is deposited on top of the device. On the hBN flake, an additional graphene flake is deposited and contacted by gold contacts (top layer). (b) False-color SFM image of the final device: the graphene top layer is colored in white and the corresponding metal contacts in red. The ribbon layer contacts as well as the overlay of the ribbon (not visible in SFM scan) are marked in blue. The orange background is the hBN flake. (c) Zoom into [b]: the top layer is not flat on a large scale but the area directly above the ribbon is wrinkle-free. The wrinkles and bubbles are highlighted with red dashed lines. (d) SFM image of the graphene nanoribbon before transfer of the hBN flake.} %
\label{fig:RibbonWithTopgate:Device}%
\end{figure}

The investigated device is shown in \Figref{RibbonWithTopgate:Device}. The single layer graphene nanoribbon was patterned by reactive ion etching, is $200\myUnit{nm}$ long and about $65\myUnit{nm}$ wide. The ribbon was fabricated on silicon dioxide since using hexagonal boron nitride (hBN) substrates complicates the fabrication while not significantly changing the electronic transport properties for such ribbons~\cite{Bischoff2012}. Two graphene side gates are located left and right of the ribbon. As visible in \Figref{RibbonWithTopgate:Device}d, there are PMMA residues on the ribbon that remained after chemical cleaning. Hereafter, the layer with the ribbon and the side gates is called \enquote{ribbon layer}. On top of the ribbon layer, a hBN flake with a thickness of $13\myUnit{nm}$ -- as determined by scanning force microscopy (SFM) -- was deposited. An additional single layer graphene flake was transferred on top of the hBN flake (top layer). The top layer flake exhibits wrinkles and bubbles and is larger than the ribbon layer graphene flake as shown in \Figref{RibbonWithTopgate:Device}b. As shown in  \Figref{RibbonWithTopgate:Device}c, the area directly above the ribbon is free of big wrinkles and bubbles. For fabrication details see Refs.~\cite{Bischoff2012,Dean2010}.

All measurements were performed at a temperature of $1.3\myUnit{K}$ and all DC voltages were applied relative to the ribbon as shown in \Figref{RibbonWithTopgate:Device}a. No leakage currents between the ribbon, the top layer or any of the gates could be measured. Unless stated differently, $V_{SG,1}=V_{SG,2}=0\myUnit{V}$.

\begin{figure}[tbp]
\centering
\includegraphics[width=\columnwidth]{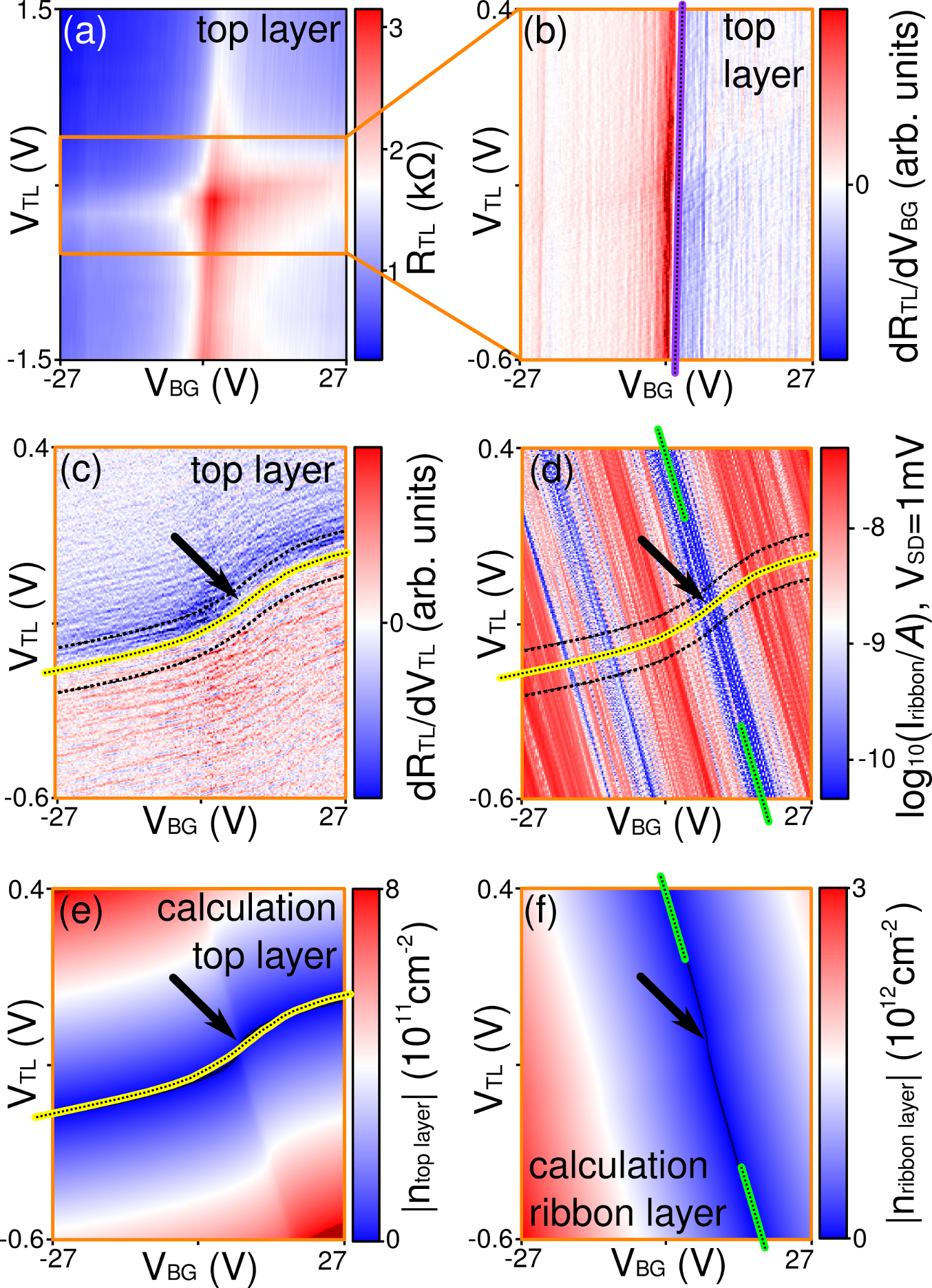} %
\caption{(color online) (a) Two-terminal resistance of the graphene top layer as a function of $V_{TL}$ and $V_{BG}$. The orange box marks the range of the following plots. (b) Derivative of [a] along $V_{BG}$. The change in sign (purple line) marks the position of the charge neutrality point of the top layer that is not above the ribbon layer. (c) Derivative of [a] along $V_{TL}$. The yellow line marks the position of the charge neutrality point in the top layer that is located above the ribbon layer. The black dashed lines mark the range where charge puddles are expected (disorder density) and the arrow marks the position where the kink of the line is the strongest. (d) Current flowing through the ribbon as a function of $V_{TL}$ and $V_{BG}$ voltage at a fixed bias voltage. The green lines mark the position of the region of suppressed conductance. (e,f) Absolute value of the calculated charge carrier density for the same range as in [c,d].}%
\label{fig:RibbonWithTopgate:AreaPlots}%
\end{figure}

\Figref{RibbonWithTopgate:AreaPlots}a shows the resistance of the graphene top layer as a function of $V_{TL}$ and $V_{BG}$ at a fixed current bias of $200\myUnit{nA}$. The voltage drop is measured over the two contacts marked in \Figref{RibbonWithTopgate:Device}b while all other top layer contacts are left floating. Contact and cable resistances of a few hundred Ohms are not subtracted. There are two pronounced features in \Figref{RibbonWithTopgate:AreaPlots}a, each being roughly parallel to one of the gate axes. \Figref{RibbonWithTopgate:AreaPlots}b shows the derivative of the resistance along $V_{BG}$. At the position where the derivative changes sign, a slope of one (purple line) is observed. This line corresponds to the charge neutrality point of the parts of the top layer graphene that are exceeding the area of the ribbon layer graphene. Therefore, the density is kept constant if $V_{TL}$ is compensated by an equal amount of $V_{BG}$. In the following, the term \enquote{top layer} will only be used for the graphene top layer part lying above the ribbon layer.  

The derivative along $V_{TL}$ is more complex as shown in \Figref{RibbonWithTopgate:AreaPlots}c: the line along which the derivative changes sign (yellow line) -- i.e. the position of the charge-neutrality point of the top layer covering the ribbon layer -- is not straight but bent. This is expected for such a system of stacked graphene layers~\cite{Kim2012} and the position of the kink (black arrow) marks the position of the average charge neutrality point of the ribbon layer. At the kink, the average charge carrier density in the ribbon layer approaches zero and screening of the back-gate is reduced. In addition to the position of the charge neutrality point of the top layer (yellow line), the outer two dashed lines mark the boundaries of the approximate region where electron-hole puddles are expected in the graphene top layer due to disorder originating from fabrication. These values are extracted at a constant $V_{BG}$. For sufficiently high density in the ribbon layer, values of $n_{dis} \approx \pm1\times 10^{11}\myUnit{cm^{-2}}$ are found for the top layer independently of the chosen $V_{BG}$~\cite{Du2008}. A density of $\pm1\times 10^{11}\myUnit{cm^{-2}}$ corresponds to a mean spacing of about $30\myUnit{nm}$ between charge carriers inside the graphene layer, and is therefore larger than the distance between the two layers ($13\myUnit{nm}$).

\Figref{RibbonWithTopgate:AreaPlots}d shows the current flowing through the ribbon as a function of $V_{TL}$ and $V_{BG}$ at a DC-bias of $1\myUnit{mV}$. Unexpectedly, there are two regions of suppressed conductance (blue, compare e.g.~\cite{Han2007,Todd2009,Molitor2009}). The right one coincides well with the bending of the yellow line (see arrow) and can therefore be interpreted as the average charge neutrality point of the ribbon layer. The origin of the left region of suppressed conductance is unclear - we believe that it might originate from inhomogeneous doping due to fabrication residues.

In Figs.~\ref{fig:RibbonWithTopgate:AreaPlots}e,f, the average densities expected in both the top ($n_T$) and ribbon layer ($n_R$) graphene are calculated from~\cite{Kim2012,Droescher2012capacitance}:
\begin{align}
V_{BG}-V_{BG,Dirac} &= \frac{e(n_R + n_T)}{C_{SiO2}} + E_F(n_R)/e \label{eq:Kim1}\\
V_{TL}-V_{TL,Dirac} &= E_F(n_R)/e - E_F(n_T)/e - \frac{e n_T}{C_{BN}}\label{eq:Kim2}\\
E_F(n) &= \frac{n}{|n|} \hbar v_F \sqrt{\pi |n|}\label{eq:Kim3}
\end{align}
$V_{TL,Dirac}=-0.05\myUnit{V}$ and $V_{BG,Dirac}=7.5\myUnit{V}$ denote the values where both the ribbon layer and the top layer are simultaneously at the charge neutrality point. $v_F=10^6\myUnit{m/s}$ is the Fermi velocity of charge carriers in graphene. The capacitances are calculated based on a parallel plate capacitor model (with $\epsilon_{SiO2}\approx 3.9$, $\epsilon_{hBN}\approx 4$, $t_{hBN}\approx 13\myUnit{nm}$, $t_{SiO2}\approx 285\myUnit{nm}$). For both layers, the evolution of the Dirac point in $V_{BG}$ and $V_{TL}$ is generally well predicted - despite the simplification of the parallel plate capacitor model which is, strictly speaking, not valid for the ribbon itself. Also the idealized $E_F(n)$ relation will be modified in experiment by disorder as will be discussed later in more detail.

\begin{figure}[tbp]
\centering
\includegraphics[width=1.0\columnwidth]{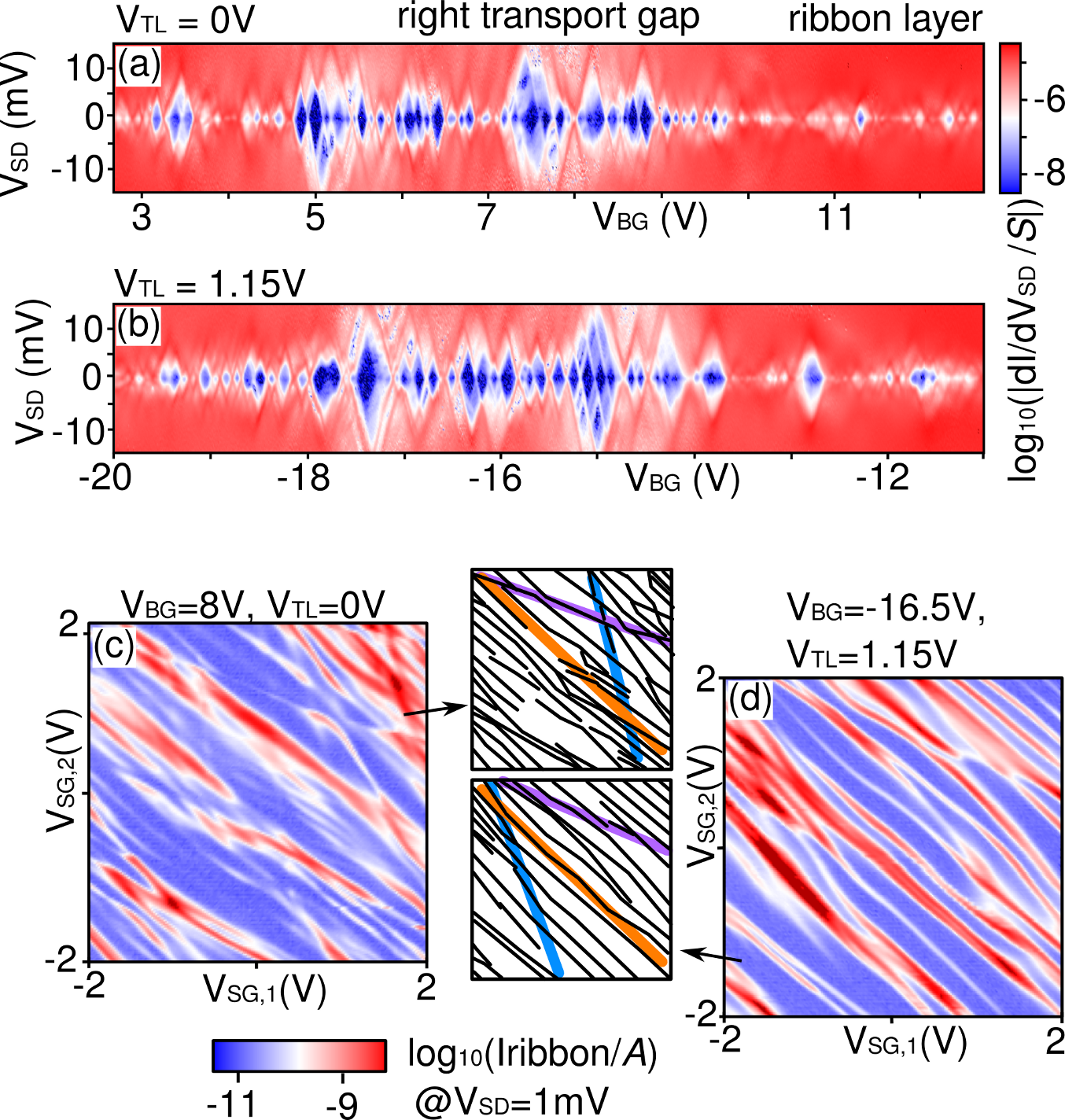} %
\caption{(color online) (a,b) Current flowing through the ribbon as a function of applied bias and $V_{BG}$ for the right region of suppressed conductance. (c,d) Current flowing through the ribbon as a function of applied side gate voltages at a fixed $V_{BG}$ and $V_{TL}$. For better visibility, the features of enhanced current are extracted and shown in the middle. (a,c) are recorded at low and (b,d) at high top layer charge carrier density.}%
\label{fig:RibbonWithTopgate:SGvsSG}%
\end{figure}

\begin{figure*}[tbp]
\centering
\includegraphics[width=\textwidth]{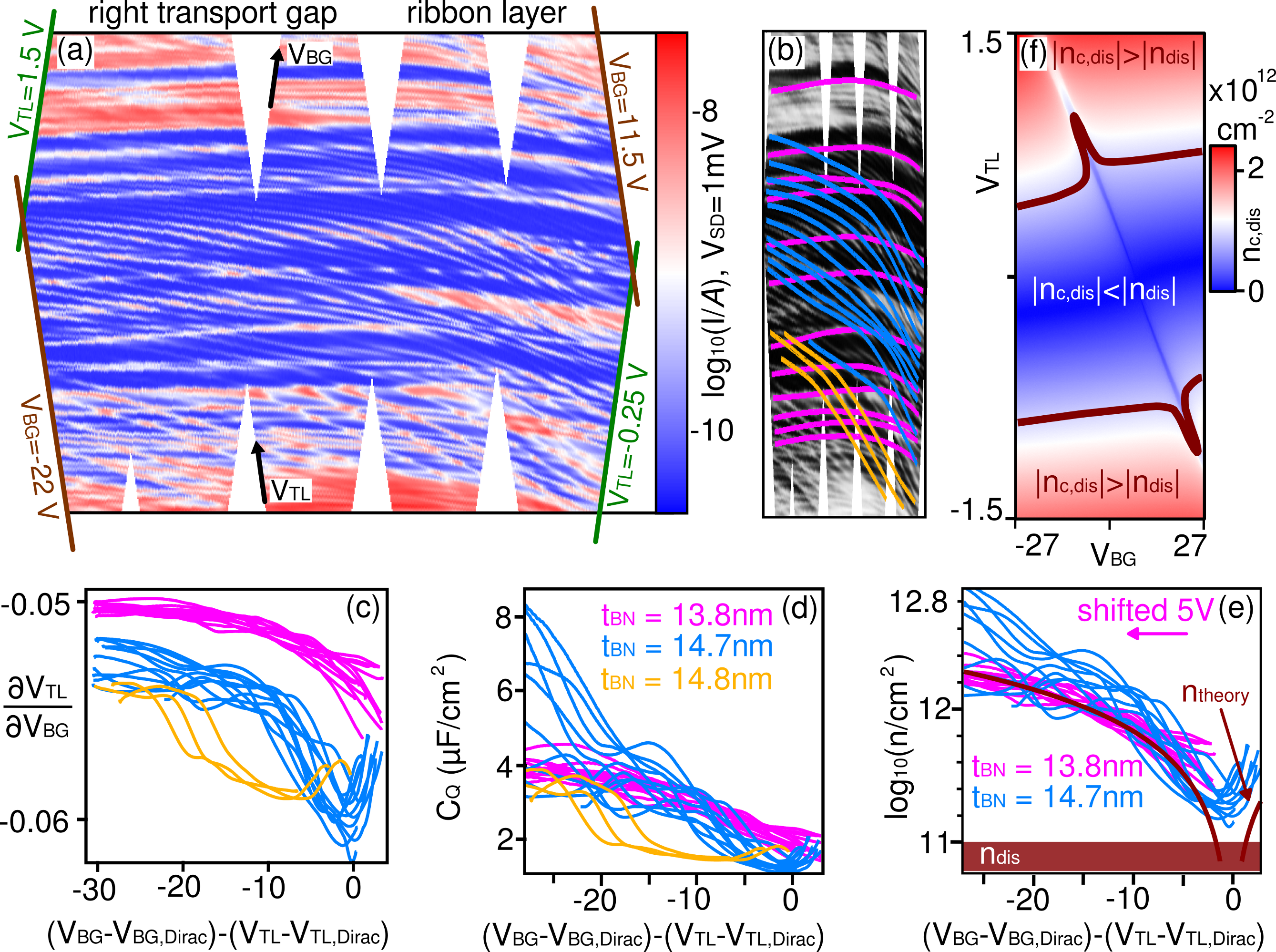} %
\caption{(color online) (a) Current flowing through the ribbon as a function of $V_{TL}$ and $V_{BG}$ at a fixed bias. In order to improve the electrical stability of the measurements, several rectangular regions in $V_{BG}$ and $V_{TL}$ were measured consecutively. For each of these measurements, the axes were transformed (rotation followed by horizontal squeeze -- similar to the n-D transformation in bilayer graphene~\cite{Weitz2010}) in order to obtain a better visibility of the features. Finally, the different measurements were put together in one plot. Lines of constant $V_{TL}$ and $V_{BG}$ are marked in green and brown. The direction of the axes $V_{TL}$ and $V_{BG}$ is marked with black arrows. (b) Same data as [a] where a number of different Coulomb resonances are marked. (c) Slopes extracted from the lines in [b] as a function of the applied gate voltages. (d) Local quantum capacitance of the top layer calculated from the slopes in [c] using different thicknesses for the hBN layer. (e) Local charge carrier density of the top layer using the data from [d] and the same hBN thicknesses as before. Magenta curves are shifted to the left. (f) Calculated \enquote{capacitance disorder} $n_{c,dis}$ with a locally varying dielectric thickness of $2\myUnit{nm}$ and a mean thickness of $14.5\myUnit{nm}$. The brown lines mark the position at which this \enquote{capacitance disorder} starts to dominate over the intrinsic fabrication disorder of around $10^{11}\myUnit{cm^{-2}}$.}%
\label{fig:RibbonWithTopgate:GapRegion}%
\end{figure*}

Next, the transport regime of suppressed conductance in the ribbon is investigated (so-called \enquote{transport gap}). Figs.~\ref{fig:RibbonWithTopgate:SGvsSG}a,b show the current flowing through the ribbon as a function of $V_{BG}$ and bias voltage for two different values of $V_{TL}$. In agreement with previous work, partially overlapping Coulomb blockade diamonds are observed~\cite{Molitor2009,Todd2009,Liu2009,Bischoff2012,Bischoff2014}, indicating the presence of multiple sites of localized charge inside the ribbon. Figs.~\ref{fig:RibbonWithTopgate:SGvsSG}c,d show the current at fixed bias flowing through the ribbon in the right transport gap as a function of the applied side gate voltages (fixed $V_{BG}$ and $V_{TL}$). While details vary, predominantly slopes of -1 are found indicating that the localized charges are equally well coupled to both side gates and therefore likely located in the ribbon. Additionally, other slopes are found indicating states that are not positioned symmetric relative to the ribbon axis and might be localized along the edges~\cite{Bischoff2014}. The observation of avoided crossings between different resonances indicates strong capacitive coupling between different sites of localized charge~\cite{Bischoff2013}. No significant difference is found when comparing transport in the right transport gap for different densities in the top layer (compare Figs.~\ref{fig:RibbonWithTopgate:SGvsSG}a,b and Figs.~\ref{fig:RibbonWithTopgate:SGvsSG}c,d). The behavior of the left transport gap is qualitatively similar to the right transport gap but current is generally less suppressed and the Coulomb blockade diamonds are smaller.

In the following, the evolution of the Coulomb resonances as a function of $V_{BG}$ and $V_{TL}$ is investigated as shown in \Figref{RibbonWithTopgate:GapRegion}a. In order to enhance the visibility of the features, the axes were transformed (details see figure caption). It is instructive to interpret the bending of the lines as the influence of the quantum capacitance rather than screening (both interpretations are equivalent). \Eqref{Kim1,Kim2,Kim3} can be rewritten using the definition of the quantum capacitance $C_Q$~\cite{Droescher2012capacitance}:
\begin{align}
C_Q(n) = e^2 D(E_F) &= \frac{e^2}{\partial E_F(n) / \partial n} \label{eq:rew1}\\ 
\left.\frac{\partial V_{TL}}{\partial V_{BG}}\right|_{n_{R}=const} &= -\frac{C_{SiO2}}{C_{BN}} - \frac{C_{SiO2}}{C_{Q}^{T}(n_T)}\label{eq:rew2}\\
\left.\frac{\partial V_{TL}}{\partial V_{BG}}\right|_{n_{T}=const} &= \frac{C_{SiO2}}{C_{SiO2}+C_Q^{R}(n_R)}\label{eq:rew3}
\end{align}
The slope of features of constant density in the ribbon layer ($\partial V_{TL}/\partial V_{BG}~|_{n_{R}=const}$) yields direct information about the quantum capacitance of the top layer ($C_{Q}^{T}(n_T)$). As Coulomb resonances in the ribbon mark the position where one electron is added to the ribbon, they are used to extract such slopes as shown in  \Figref{RibbonWithTopgate:GapRegion}b. \Figref{RibbonWithTopgate:GapRegion}c shows the value of the slopes as a function of gate voltages. It is difficult to exactly track the resonances due to avoided crossings between different resonances lead to small wiggles in the extracted slopes. There are at least 3 sets of slopes that are marked in different colors. Due to the localization of charge carriers in the ribbon, the Coulomb resonances are a local sensor for the electronic properties of the top layer. Also note that \Eqref{rew1,rew2,rew3} are generic and do not contain graphene specific terms.

We now use the evolution of the Coulomb blockade resonances to obtain information on the local quantum capacitance of the top layer. In the limit of high top layer density (left side of \Figref{RibbonWithTopgate:GapRegion}c, saturation of slope), the quantum capacitance is negligible and the slope is given by $-C_{SiO2}/C_{BN}$. We find a different saturation value of $\partial V_{TL} / \partial V_{BG}$ for different sets of lines. While it is possible that different sites of localized charge experience a different capacitance ratio due to geometry, we believe it to be more likely that the observed differences originate from a local change in $C_{BN}$ due to fabrication residues that locally increase the thickness of the dielectric. 

Aside from a different saturation value, the different curves are shifted horizontally, indicating a different local doping of the top layer graphene that is probed by the localized state in the ribbon. The shift between the three different sets of lines can be explained by different induced densities due to the different local capacitances together with an nonzero intrinsic doping of the top layer graphene. 

As found in previous experiments, localized states can slightly change their spatial position when an additional electron is added or when the electrostatic environment is changed~\cite{Pascher2012,Bischoff2014}. The deviations within one set of lines can therefore be explained by slightly different areas of the top layer that are probed by the localized states in the ribbon.

In \Figref{RibbonWithTopgate:GapRegion}d, the extracted slopes are converted into a local quantum capacitance of the top layer using \Eqref{rew2} and different hBN thicknesses based on the saturation value of the slope. The quantum capacitance is converted into a local charge carrier density using \Eqref{rew1} and by shifting the magenta curves by $5\myUnit{V}$ to the left as shown in \Figref{RibbonWithTopgate:GapRegion}e (the yellow lines are not shown for clarity). The thick brown line ($n_{theory}$) is calculated using \Eqref{rew1} together with the assumption that the middle layer is at zero density. 

Generally, the lines extracted from the measurement fit well to the expected values. The blue lines have a much wider spread around the theoretical value than the red lines. This is compatible with the model of a spatially changing capacitance between the ribbon and the top layer due to fabrication residues: the red lines correspond to areas that are mostly flat, whereas the blue lines correspond to areas with residues. As the areas on which the charges are localized in the ribbon are likely larger than the typical area of fabrication residues~\cite{Bischoff2012,Bischoff2014}, spatial averaging of the capacitance occurs. This is compatible with the fact that the extracted value from the measurement ($\approx 1\myUnit{nm}$) is smaller than the value from the SFM scan ($5\myUnit{nm}$). The shift of the magenta lines indicates a local difference in doping.

In order to estimate in which regime this local change of capacitance is important, \Eqref{Kim1,Kim2,Kim3} are used to calculate the difference in density ($n_{c,dis}$) if the thickness of the hBN flake is varied locally by $2\myUnit{nm}$ as shown in \Figref{RibbonWithTopgate:GapRegion}f. Due to the plate capacitor geometry, $n_{c,dis}$ is equally large for both ribbon and top layer. As soon as the brown lines are crossed ($n_{c,dis}=n_{dis}$), the effect of varying dielectric thickness starts to dominate over the intrinsic doping fluctuations due to fabrication residues. These conclusions are important, for example, for double-gated bilayer graphene devices where a high top-gate voltage is applied in order to open a bandgap~\cite{Oostinga2007,Allen2012,Droescher2012,Goosens2012,Velasco2012,Young2012,Varlet2014}. Choosing a thick dielectric together with as clean interfaces as possible mitigates these layer thickness fluctuations.

Finally it is possible to extract the average quantum capacitance of the ribbon layer using the data from \Figref{RibbonWithTopgate:AreaPlots}a together with \Eqref{rew3} and tracking the position of the Dirac point in the top layer. The extracted density as a function of the gate voltages fits reasonably well to the expectations and a disorder density $n_{dis,R}\approx 1-2\times 10^{11}\myUnit{cm^{-2}}$ is obtained for the ribbon layer.

In summary, we have investigated electrical transport through a graphene nanoribbon and a strongly capacitively coupled graphene top layer. As this experiment is a crucial step towards stacked -- and therefore strongly coupled -- graphene nanostructures, it is important to understand transport in this system in detail. We showed that the mutual interactions can be discussed in terms of screening or, equivalently, quantum capacitance. In agreement with previous studies, transport through the nanoribbon was governed by Coulomb blockade in the regime of low density. The evolution of single Coulomb blockade peaks was tracked as a function of voltages applied to the back gate and the top layer graphene. It was shown that the position of the Coulomb resonances in gate voltage space is directly related to the quantum capacitance and therefore, the density of the top layer graphene sheet. A detailed analysis of the different extracted Coulomb peak positions showed different geometrical capacitances per unit area between localized charges in the ribbon and the top layer graphene. This behaviour can be explained with fabrication residues changing the thickness of the dielectric locally. These findings show experimentally that it is crucial for stacked graphene devices to have interfaces that are as  clean as possible. Small variations in dielectric thickness can quickly become more important to transport than the intrinsic electron-hole puddles originating from nonzero chemical doping of the devices. Finally, it is worth noting that charge localization inside the graphene nanoribbon is not influenced strongly by the density of the top layer graphene, except for the mentioned evolution in gate voltage.

\section*{Acknowledgements}
We thank Florian Libisch for helpful discussions. Financial support by the National Center of Competence in Research on “Quantum Science and Technology“ (NCCR QSIT) funded by the Swiss National Science Foundation is gratefully acknowledged.




\begin{thebibliography}{99}

\bibitem{Georgiou2012}
T. Georgiou, R. Jalil, B. D. Belle, L. Britnell, R. V. Gorbachev, S. V. Morozov, Y.-J. Kim, A.  Gholinia, S. J. Haigh, O. Makarovsky, L. Eaves, L. A. Ponomarenko, A. K. Geim, K. S. Novoselov, A. Mishchenko, Nat. Nano. {\bf 8}, 100-103 (2012).

\bibitem{Britnell2012science}
L. Britnell, R. V. Gorbachev, R. Jalil, B. D. Belle, F. Schedin, A. Mishchenko, T. Georgiou, M. I. Katsnelson, L. Eaves, S. V. Morozov, N. M. R. Peres, J. Leist, A. K. Geim, K. S. Novoselov, L. A. Ponomarenko, Science {\bf 335}, 947-950 (2012).

\bibitem{Britnell2012nano}
L. Britnell, R. V. Gorbachev, R. Jalil, B. D. Belle, F. Schedin, M. I. Katsnelson, L. Eaves, S. V. Morozov, A. S. Mayorov, N. M. R. Peres, A. H. Castro Neto, J. Leist, A. K. Geim, L. A. Ponomarenko, K. S. Novoselov, Nanoletters {\bf 12}, 1707-1710 (2012).

\bibitem{Yang2012}
H. Yang, J. Heo, S. Park, H. J. Song, D. H. Seo, K.-E. Byun, P. Kim, IK. Yoo, H.-J. Chung, K. Kim, Science {\bf 336}, 1140-1143 (2012).

\bibitem{Lee2013}
G.-H. Lee, Y.-J. Yu, X. Cui, N. Petrone, C.-H. Lee, M. S. Choi, D.-Y. Lee, C. Lee, W. J. Yoo, K. Watanabe, T. Taniguchi, C. Nuckolls, P. Kim, J. Hone, ACS Nano {\bf 7}, 7931-7936 (2013).

\bibitem{Lee2014}
C.-H. Lee, G.-H. Lee, A. M. van der Zande, W. Chen, Y. Li, M. Han, X. Cui, G. Arefe, C. Nuckolls, T. F. Heinz, J. Guo, J. Hone, P. Kim, Nature Nano. {\bf 9}, 676-681 (2014).

\bibitem{Choi2013}
M. S. Choi, G.-H. Lee, Y.-J. Yu, D.-Y. Lee, S. H. Lee, P. Kim, J. Hone, W. J. Yoo, Nat. Comm. {\bf 4}, 1624 (2013).

\bibitem{Liu2012}
M. Liu, X. Yin, X. Zhang, Nanoletters {\bf 12}, 1482-1485 (2012).

\bibitem{Ponomarenko2011}
L. A. Ponomarenko, A. K. Geim, A. A. Zhukov, R. Jalil, S. V. Morozov, K. S. Novoselov, I. V. Grigorieva, E. H. Hill, V. V. Cheianov, V. I. Fal'ko, K. Watanabe, T. Taniguchi, R. V. Gorbachev, Nat. Phys. {\bf 7}, 958-961 (2011).

\bibitem{Kechedzhi2012}
K. Kechedzhi, E. H. Hwang, S. Das Sarma, PRB {\bf 86}, 165442 (2012).

\bibitem{Kim2012}
S. Kim, E. Tutuc, Solid State Comm {\bf 152}, 1283-1288 (2012).

\bibitem{Gorbachev2012}
R. V. Gorbachev, A. K. Geim, M. I. Katsnelson, K. S. Novoselov, T. Tudorovskiy, I. V. Grigorieva, A. H. MacDonald, S. V. Morozov, K. Watanabe, T. Taniguchi, L. A. Ponomarenko, Nat. Phys. {\bf 8}, 896-901 (2012).

\bibitem{Titov2013}
M. Titov, R. V. Gorbachev, B. N. Narozhny, T. Tudorovskiy, M. Schutt, P. M. Ostrovsky,I. V. Gornyi, A. D. Mirlin, M. I. Katsnelson, K. S. Novoselov, A. K. Geim, L. A. Ponomarenko, PRL {\bf 111}, 166601 (2013).

\bibitem{Yankowitz2012}
M. Yankowitz, J. Xue, D. Cormode, J. D. Sanchez-Yamagishi, K. Watanabe, T. Taniguchi, P. Jarillo-Herrero, P. Jacquod, B. J. LeRoy, Nat. Phys. {\bf 8}, 382-386 (2012).

\bibitem{Dean2013}
C. R. Dean, L. Wang, P. Maher, C. Forsythe, F. Ghahari, Y. Gao, J. Katoch, M. Ishigami, P. Moon, M. Koshino, T. Taniguchi, K. Watanabe, K. L. Shepard, J. Hone, P. Kim, Nature {\bf 497}, 598-602 (2013).

\bibitem{Ponomarenko2013}
L. A. Ponomarenko, R. V. Gorbachev, G. L. Yu, D. C. Elias, R. Jalil, A. A. Patel, A. Mishchenko, A. S. Mayorov, C. R. Woods, J. R. Wallbank, M. Mucha-Kruczynski, B. A. Piot, M. Potemski, I. V. Grigorieva, K. S. Novoselov, F. Guinea, V. I. Fal'ko, A. K. Geim, Nature {\bf 497}, 594-597 (2013).

\bibitem{Molitor2009}
F. Molitor, A. Jacobsen, C. Stampfer, J. G\"uttinger, T. Ihn, K. Ensslin, PRB {\bf 79}, 075426 (2009).

\bibitem{Todd2009}
K. Todd, H.-T. Chou, S. Amasha, D. Goldhaber-Gordon, Nano Lett. {\it 9}, 416-421 (2009).

\bibitem{Liu2009}
X. Liu, J. B. Oostinga, A. F. Morpurgo, L. M. K. Vandersypen, PRB {\bf 80}, 121407 (2009).
 
\bibitem{Han2007}
M. Y. Han, B. \"Ozyilmaz, Y. Zhang, P. Kim, PRL {\bf 98}, 206805 (2007).

\bibitem{Bischoff2012}
D. Bischoff, T. Krähenmann, S. Dröscher, M. A. Gruner, C. Barraud, T. Ihn, K. Ensslin, APL {\bf 101}, 203103 (2012).

\bibitem{Wei1998}
Y. Y. Wei, J. Weis, K. v. Klitzing, K. Eberl, PRL {\bf 81}, 1674-1677 (1998).

\bibitem{Hess1998}
H. F. Hess, T. A. Fulton, M. J. Yoo, A. Yacoby, Solid State Communications {\bf 107}, 657-661 (1998).

\bibitem{Martin2009}
J. Martin, N. Akerman, G. Ulbricht, T. Lohmann, K. von Klitzing, J. H. Smet, A. Yacoby, Nat Phys {\bf 5}, 669-674 (2009).

\bibitem{Dean2010}
C. R. Dean, A. F. Young, I. Meric, C. Lee, L. Wang, S. Sorgenfrei, K. Watanabe, T. Taniguchi, P. Kim, K. L. Shepard, J. Hone, Nature Nano. {\bf 5}, 722 (2010).

\bibitem{Du2008}
X. Du, I. Skachko, A. Barker, E. Y. Andrei, Nat Nano {\bf 3}, 491-495 (2008).

\bibitem{Droescher2012capacitance}
S. Dröscher, P. Roulleau, F. Molitor, P. Studerus, C. Stampfer, K. Ensslin, T. Ihn, Physica Scripta {\bf T146}, 014009 (2012).

\bibitem{Bischoff2014}
D. Bischoff, F. Libisch, J. Burgd\"orfer, T. Ihn, K. Ensslin, PRB {\bf 90}, 115405 (2014).

\bibitem{Bischoff2013}
D. Bischoff, A. Varlet, P. Simonet, T. Ihn, K. Ensslin, New Jour. of Phys {\bf 15}, 083029 (2013).

\bibitem{Weitz2010}
R. T. Weitz, M. T. Allen, B. E. Feldman, J. Martin, A. Yacoby, Science {\bf 330}, 812-816 (2010).

\bibitem{Pascher2012}
N. Pascher, D. Bischoff, T. Ihn, K. Ensslin, APL {\bf 101}, 063101 (2012).

\bibitem{Oostinga2007}
J. B. Oostinga, H. B. Heersche, X. Liu, A. F. Morpurgu, L. M. K. Vandersypen, Nature Mat. {\bf 7}, 151-157 (2007).

\bibitem{Allen2012}
M. T. Allen, J. Martin, A. Yacoby, Nat. Com. {\bf 3}, 934 (2012).

\bibitem{Droescher2012}
S. Dröscher, C. Barraud, K. Watanabe, T. Taniguchi, T. Ihn, Klaus Ensslin, NJP {\bf 14}, 103007 (2012).

\bibitem{Goosens2012}
A. M. Goossens, S. C. M. Driessen, T. A. Baart, K. Watanabe, T. Taniguchi, L. M. K. Vandersypen, Nano Lett. {\bf 12}, 4656 (2012).

\bibitem{Velasco2012}
J. Velasco Jr., L. Jing, W. Bao, Y. Lee, P. Kratz, V. Aji, M. Bockrath, C. N. Lau1, C. Varma, R. Stillwell, D. Smirnov, F. Zhang, J. Jung, A. H. MacDonald, Nature Nano {\bf 7}, 156-160 (2012).

\bibitem{Young2012}
A. F. Young, C. R. Dean, I. Meric, S. Sorgenfrei, H. Ren, K. Watanabe, T. Taniguchi, J. Hone, K. L. Shepard, P. Kim, PRB {\bf 85}, 235458 (2012).

\bibitem{Varlet2014}
A. Varlet, D. Bischoff, P. Simonet, K. Watanabe, T. Taniguchi, T. Ihn, K. Ensslin, M. Mucha-Kruczynski, V. I. Fal'ko, PRL {\bf 113}, 116602 (2014).


\end{thebibliography}
\end{document}